# Simple derivation of the Generalized Möller-Wu-Lee transformations. Born rigid constant accelerated motion on a curved Lorentzian manifold.

Jaykov Foukzon

Israel Institute of Technology, Haifa, Israel

S.A.Podosenov

All-Russian Scientific-Research Institute
for Optical and Physical Measurements,
Moscow 119361, Russia

**Abstract.** Simple derivation of the classical generalized Möller-Wu-Lee transformations from general master equation is presented. We will argue that in fact we can implement Born's notion of rigid motion in both flat spacetime and arbitrary curved non-holonomic spacetimes containing classical and Colombeau's distributional sources.

## I. Introduction

Hsu and Kleff [1], [2] derived a nonlinear coordinate transformation using approach, which is based on earlier works by Möller [3], Wu, and Lee [4], the generalized Möller-Wu-Lee (MWL) transformations:

$$ct = \frac{c^2}{2\alpha\gamma_0} \ln\left[\frac{(1 + \alpha x_I/c^2 + \alpha t_I/c + \beta_0)(1 - \beta_0)}{(1 + \alpha x_I/c^2 - \alpha t_I/c - \beta_0)(1 + \beta_0)}\right],$$

$$x = \left(\frac{c^2}{\alpha}\right)\left\{\left[\left(1 + \frac{\alpha x_I}{c^2}\right)^2 - \left(\frac{\alpha t_I}{c} + \beta_0\right)^2\right]^{1/2} - \frac{1}{\gamma_0}\right\}$$

$$y_I = y, z_I = z,$$

$$\beta_0 = \frac{v}{c}, \gamma_0 = \frac{1}{\sqrt{1 - \beta_0^2}}.$$

(1.1)

The inverse of the generalized MWL transformation (1.1) is given by:

$$ct_I = \gamma_0\left(x + \frac{c^2}{\alpha\gamma_0}\right)\sinh\left[\left(\frac{t\alpha\gamma_0}{c}\right) + \tanh^{-1}\beta_0\right] - \frac{\beta_0 c^2}{\alpha},$$

$$x_I = \left(x + \frac{c^2}{\alpha\gamma_0}\right)\cosh\left[\left(\frac{t\alpha\gamma_0}{c}\right) + \tanh^{-1}\beta_0\right] - \frac{c^2}{\alpha}, \qquad (1.2)$$

$$y_I = y, z_I = z.$$

Ernst [5] derived spacetime transformations involving constant linear accelerations on the basis of works by Misner, Thorne and Wheeler on hyperbolic motion and the principle of limiting four-dimensional symmetry by Hsu. The transformations:

$$ct_I = \gamma_0\left(x + \frac{c^2}{\alpha}\right)\sinh\left[\left(\frac{t\alpha}{c}\right) + \tanh^{-1}\beta_0\right] - \frac{\beta_0\gamma_0 c^2}{\alpha},$$

$$x_I = \left(x + \frac{c^2}{\alpha}\right)\cosh\left[\left(\frac{t\alpha}{c}\right) + \tanh^{-1}\beta_0\right] - \frac{\gamma_0 c^2}{\alpha}, \qquad (1.3)$$

$$y_I = y, z_I = z.$$

The inverse **MWL** transformations (1.2) and the Ernst transformation (1.3) are related by a change of variables $\alpha \leftrightarrow \alpha\gamma_0$.

In [1] and [5] the formulas (1.1)-(1.3) were obtained in a very long and complicated way by solving a system of a differential equations.

**Remark**   1.1. We obtain the inverse MWL transformations from the general master equation (1.5) by a very short and simple way.

**Definition**   1.1. The metric $ds_W^2 = g_{\mu\nu}(x^\nu)dx^\mu dx^\nu$ is forming by nonlinear transformations $x_I^\nu = W^\nu(x^\nu); W : F \to F_I = M^4$ iff

$$g_{\mu\nu}(x^\nu) = \sum_\alpha \frac{\partial x_I^\alpha}{\partial x^\mu}\frac{\partial x_I^\alpha}{\partial x^\nu} = \sum_\alpha \frac{\partial W^\alpha(x^\nu)}{\partial x^\mu}\frac{\partial W^\alpha(x^\nu)}{\partial x^\nu},$$

$$(1.4)$$

$$\alpha, \mu, \nu = 0, 1, 2, 3$$

**Theorem**   1.1. Suppose:
1. A metric $ds_W^2$ is forming by nonlinear transformations $W : F \to F_I$ where $x_I^\nu = W^\nu(x^\nu)$;
2. A metric $ds_W^2[v]$ is forming by nonlinear transformations $W_{L_i^\nu} : F \to F_I$ where $x_I^\nu = W_{L_i^\nu}$;

$$W_{L_i^\nu} = L_i^\nu[W(x^\nu)], \nu = 0, 1, 2, 3 \qquad (1.5)$$

*where* $(L_i^v)^1 = \gamma(v)(x^1 + \beta x^0), (L_i^v)^0 = \gamma(v)(x^0 + \beta x^1), \beta = \frac{v}{c}$.
*Then:*

$$ds_W^2 = ds_W^2[v],$$

$$[ds_W^2]_{x^v = W_{L_i^v}^{-1}(x_I^v)} = c^2 dt_I^2 - (dx_I^1)^2 - (dx_I^2)^2 - (dx_I^3)^2$$

(1.6)

*where the subscript* $[\circ]_{x^v = W_{L_i^v}^{-1}(x_I^v)}$ *denotes the substitution*
$x^v = W_{L_i^v}^{-1}(x_I^v), v = 0, 1, 2, 3.$

## II. Möller accelerated observers

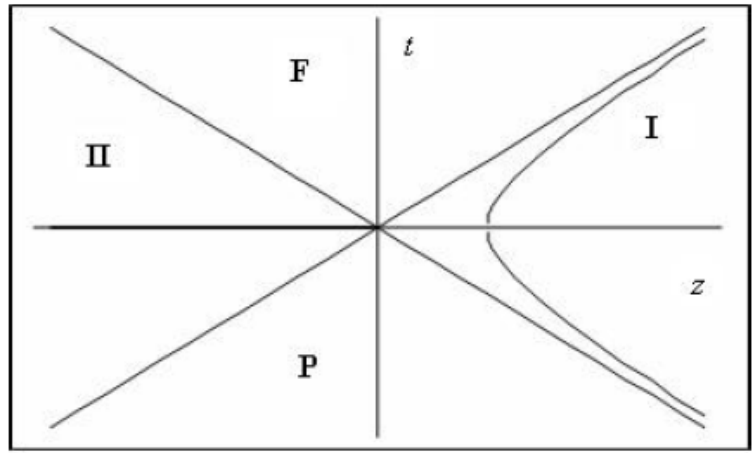

**Fig. 1.** The division of Minkowski space into four regions,
the right Rindler wedge (**I**), the left Rindler wedge (**II**),
future (**F**), and past (**P**). The hyperbola is the worldline of
an accelerated observer.

We consider the case of a uniformly accelerated observer in Minkowski space-time $M^4$ as viewed from an inertial reference frame $F_I$. As shown in Fig.1, we consider Minkowski space with coordinates $(z_I^0, z_I^1)$ divided into four regions **I,II,F,P**.

These are the right Rindler wedge (region **I**), the left Rindler wedge (region **II**) and the future (**F**) and past (**P**). We define Rindler coordinates $(v, u)$ which can be related to Minkowski coordinates $(ct_I, z_I^1) = (z_I^0, z_I^1)$ in the following way:

(a) in regions **I,II:**

$$ct_I = x_I^0 = u\sinh v, z_I^1 = u\cosh v; \qquad (a)$$

$$v = \tanh^{-1}\left(\frac{z_I^0}{z_I^1}\right), u = sgn(z_I^1)\sqrt{(z_I^1)^2 - (z_I^0)^2}. \quad (b)$$

(2.1)

(b) in regions **F,P**:

$$ct_I = z_I^0 = u\cosh v, z_I^1 = u\sinh v; \qquad (a)$$

$$v = \tanh^{-1}\left(\frac{z_I^1}{z_I^0}\right), u = sgn(z_I^0)\sqrt{(z_I^0)^2 - (z_I^1)^2}. \quad (b)$$

(2.2)

Since the calculations described in this paper will be similar for all four regions, we focus on region **I**, which we denote as Rindler space. In this region, starting with the two-dimensional Minkowski line element $ds^2 = c^2 dt^2 - dz^2$ and using (2.1.a), it is easy to show that the line element takes the form

$$ds^2 = u^2 dv^2 - du^2 \tag{2.3}$$

Note that $v$ is a timelike coordinate and $u$ is a spacelike coordinate-in region **I**. This will change in the other regions. Using this line element we can read off the components of the metric tensor. These can be arranged in a matrix as follows

$$g_{\mu\eta} = \begin{pmatrix} u & 0 \\ 0 & -1 \end{pmatrix} \tag{2.4}$$

The components of $g^{\mu\eta}$ can be found easily by inverting (2.4).

Setting $z_I^1 = x_I^1 + c^2/a, u = \xi + \frac{c^2}{a}, v = \frac{a\tau}{c}$ in regions **I,II** and setting $z_I^0 = x_I^0 + c^2/a, u = \xi + \frac{c^2}{a}, v = \frac{a\tau}{c}$ in regions **F,P** we define Möller coordinates $(\tau, \xi)$ which can be related to Minkowski coordinates $(ct_I, x_I^1) = (x_I^0, x_I^1)$ in the following way:

(a) in regions **I,II**:

$$ct_I = x_I^0 = x_I^0(\tau, \xi) = \left(\frac{c^2 + \xi a}{a}\right)\sinh\left(\frac{a\tau}{c}\right),$$

$$x_I^1 = x_I^1(\tau, \xi) = \left(\frac{c^2 + \xi a}{a}\right)\cosh\left(\frac{a\tau}{c}\right) - \frac{c^2}{a}.$$

(2.5)

(b) in regions **F,P**:

$$ct_I = x_I^0 = x_I^0(\tau,\xi) = \left(\frac{c^2+\xi a}{a}\right)\cosh\left(\frac{a\tau}{c}\right) - \frac{c^2}{a},$$

$$x_I^1 = x_I^1(\tau,\xi) = \left(\frac{c^2+\xi a}{a}\right)\sinh\left(\frac{a\tau}{c}\right). \tag{2.6}$$

The coordinates $(\tau,\xi)$ are defined in the intervals $-\infty < \tau < \infty$ and $-\frac{c^2}{a} < \xi < \infty$ and it is easily seen that the line $\xi = -\frac{c^2}{a}$ is a horizon line for the accelerated observers. A simple calculation shows that the metric of Minkowski spacetime

$$ds^2 = c^2 dt_I^2 - d(x_I^1)^2 = (dx_I^0)^2 - (dx_I^1)^2 \tag{2.7}$$

in the proper coordinates $(\xi,\tau)$ is given by

$$ds_a^2 = c^2 dt^2 - dx^2 = \left(c + \frac{\xi a}{c}\right)^2 d\tau^2 - d\xi^2. \tag{2.8}$$

From Eqs.(2.1)-(2.2) the inverse transformation of (2.5)-(2.6) is given by
  (**a**) in regions **I,II**:

$$\xi(x^0,x^1) = \text{sgn}\left(x^1 + \frac{c^2}{a}\right)\sqrt{\left(x^1+\frac{c^2}{a}\right)^2 - (x^0)^2} - \frac{c^2}{a} =$$

$$= \text{sgn}\left(x^1 + \frac{c^2}{a}\right)\frac{c^2}{a}\left[\sqrt{\left(1+\frac{ax^1}{c^2}\right)^2 - \left(\frac{ax^0}{c^2}\right)^2} - 1\right]$$

$$c\tau(x^0,x^1) = \frac{c^2}{a}\tan^{-1}\left(\frac{x^0}{x^1+\frac{c^2}{a}}\right) = \tag{2.9}$$

$$= \frac{c^2}{2a}\ln\left(\frac{\frac{c^2}{a}+x^1+x^0}{\frac{c^2}{a}+x^1-x^0}\right) = \frac{c^2}{2a}\ln\left(\frac{1+\frac{ax^1}{c^2}+\frac{ax^0}{c^2}}{1+\frac{ax^1}{c^2}-\frac{ax^0}{c^2}}\right).$$

  (**b**) in regions **F,P**:

$$\xi(x^0,x^1) = sgn\left(x^0 + \frac{c^2}{a}\right)\sqrt{\left(x^0 + \frac{c^2}{a}\right)^2 - (x^1)^2} - \frac{c^2}{a},$$

$$c\tau(x^0,x^1) = \frac{c^2}{a}\tan^{-1}\left(\frac{x^1}{x^0 + \frac{c^2}{a}}\right) = \quad (2.10)$$

$$\frac{c^2}{2a}\ln\left(\frac{\frac{c^2}{a} + x^1 + x^0}{\frac{c^2}{a} - x^1 + x^0}\right) = \frac{c^2}{2a}\ln\left(\frac{1 + \frac{ax^1}{c^2} + \frac{ax^0}{c^2}}{1 - \frac{ax^1}{c^2} + \frac{ax^0}{c^2}}\right).$$

## III. Simple derivation of the Generalized Möller-Wu-Lee (Hsu-Kleff-Ernst) transformation by general master equation

From general master equation (1.5) the inverse of the generalized Möller-Wu-Lee transformation is given by the relation:

$$\{x_I^0, x_I^1\} = L_i^\mathbf{v}[\{x_I^0(\tau,\xi), x_I^1(\tau,\xi)\}]_{\alpha=\alpha\gamma_0}. \quad (3.1)$$

where a vector function $\{x_I^0(\tau,\xi), x_I^1(\tau,\xi)\}$ is given by Eqs.(2.5)-(2.6) and where the subscript $[\circ]_{\alpha=\alpha\gamma_0}$ denotes the substitution $\alpha = \alpha\gamma_0$.

Let us calculate a vector function $\{[(L_i^\mathbf{v})^0](\tau,\xi), [(L_i^\mathbf{v})^1](\tau,\xi)\}$:

$$[(L_i^\mathbf{v})^0](\tau,\xi) = L_i^0[\{x_I^0(\tau,\xi), x_I^1(\tau,\xi)\}] = \gamma_0(\mathbf{v})[x_I^0(\tau,\xi) + \beta_0 x_I^1(\tau,\xi)],$$

(3.2)

$$[(L_i^\mathbf{v})^1](\tau,\xi) = L_i^1[\{x_I^0(\tau,\xi), x_I^1(\tau,\xi)\}] = \gamma_0(\mathbf{v})[x_I^1(\tau,\xi) + \beta_0 x_I^0(\tau,\xi)].$$

From Eqs. (2.5),(2.6),(3.2) we find the Ernst transformations [5]:

(**a**) in regions **I,II:**

$$x_I^0 = L_i^0[\{x_I^0(\tau,\xi), x_I^1(\tau,\xi)\}] =$$

$$\gamma_0\left(\frac{c^2+\xi a}{a}\right)\left[\sinh\left(\frac{a\tau}{c}\right)+\beta_0\cosh\left(\frac{a\tau}{c}\right)\right]-\frac{\gamma_0\beta_0 c^2}{a},$$

$$x_I^1 = L_i^1[\{x_I^0(\tau,\xi), x_I^1(\tau,\xi)\}] =$$

$$\gamma_0\left(\frac{c^2+\xi a}{a}\right)\left[\beta_0\sinh\left(\frac{a\tau}{c}\right)+\cosh\left(\frac{a\tau}{c}\right)\right]-\frac{\gamma_0 c^2}{a}.$$

(3.3)

(**b**) in regions **F,P**:

$$x_I^0 = L_i^0[\{x_I^0(\tau,\xi), x_I^1(\tau,\xi)\}] =$$

$$\gamma_0\left(\frac{c^2+\xi a}{a}\right)\left[\cosh\left(\frac{a\tau}{c}\right)+\beta_0\sinh\left(\frac{a\tau}{c}\right)\right]-\frac{\gamma_0 c^2}{a},$$

$$x_I^1 = L_i^1[\{x_I^0(\tau,\xi), x_I^1(\tau,\xi)\}] =$$

$$\gamma_0\left(\frac{c^2+\xi a}{a}\right)\left[\beta_0\cosh\left(\frac{a\tau}{c}\right)+\sinh\left(\frac{a\tau}{c}\right)\right]-\frac{\gamma_0\beta_0 c^2}{a}.$$

(3.4)

From replacement $\alpha \to \gamma_0\alpha$ into Eqs. (3.3),(3.4) we obtain canonical form of the Möller-Wu-Lee transformation [1]:

(**a**) in regions **I,II**:

$$x_I^0(\tau,\xi) = \gamma_0\left(\xi+\frac{c^2}{\gamma_0\alpha}\right)\left[\sinh\left(\frac{\gamma_0\alpha\tau}{c}\right)+\beta_0\cosh\left(\frac{\gamma_0\alpha\tau}{c}\right)\right]-\frac{\beta_0 c^2}{\alpha} =$$

$$= \left(\gamma_0\xi+\frac{c^2}{\alpha}\right)\left[\sinh\left(\frac{\gamma_0\alpha\tau}{c}\right)+\beta_0\cosh\left(\frac{\gamma_0\alpha\tau}{c}\right)\right]-\frac{\beta_0 c^2}{\alpha},$$

(3.5)

$$x_I^1(\tau,\xi) = \gamma_0\left(\xi+\frac{c^2}{\gamma_0\alpha}\right)\left[\beta_0\sinh\left(\frac{\gamma_0\alpha\tau}{c}\right)+\cosh\left(\frac{\gamma_0\alpha\tau}{c}\right)\right]-\frac{c^2}{\alpha} =$$

$$= \left(\gamma_0\xi+\frac{c^2}{\alpha}\right)\left[\beta_0\sinh\left(\frac{\gamma_0\alpha\tau}{c}\right)+\cosh\left(\frac{\gamma_0\alpha\tau}{c}\right)\right]-\frac{c^2}{\alpha}.$$

(**b**) in regions **F,P**:

$$x_I^0(\tau,\xi) =$$

$$\gamma_0\left(\xi + \frac{c^2}{\gamma_0\alpha}\right)\left[\cosh\left(\frac{\gamma_0\alpha\tau}{c}\right) + \beta_0\sinh\left(\frac{\gamma_0\alpha\tau}{c}\right)\right] - \frac{c^2}{\alpha},$$

(3.6)

$$x_I^1(\tau,\xi) =$$

$$\gamma_0\left(\xi + \frac{c^2}{\gamma_0\alpha}\right)\left[\beta_0\cosh\left(\frac{\gamma_0\alpha\tau}{c}\right) + \sinh\left(\frac{\gamma_0\alpha\tau}{c}\right)\right] - \frac{\beta_0 c^2}{\alpha}.$$

The inverse transformation of the Ernst transformation (3.3)-(3.4) is given by

(**a**) in regions **I,II**:

$$x^0 = \frac{c^2}{\alpha}\left[\tanh^{-1}\left(\frac{x_I^0 + \frac{\beta_0\gamma_0 c^2}{\alpha}}{x_I^1 + \frac{\gamma_0 c^2}{\alpha}}\right) - \tanh^{-1}(\beta_0)\right],$$

(3.7)

$$x^1 = \sqrt{\left(x_I^1 + \frac{\gamma_0 c^2}{\alpha}\right) - \left(x_I^0 + \frac{\beta_0\gamma_0 c^2}{\alpha}\right)} - \frac{c^2}{\alpha}$$

(**b**) in regions **F,P**:

$$x^0 = \frac{c^2}{\alpha}\left[\tanh^{-1}\left(\frac{x_I^1 + \frac{\beta_0\gamma_0 c^2}{\alpha}}{x_I^0 + \frac{\gamma_0 c^2}{\alpha}}\right) - \tanh^{-1}(\beta_0)\right],$$

(3.8)

$$x^1 = \sqrt{\left(x_I^1 + \frac{\beta_0\gamma_0 c^2}{\alpha}\right) - \left(x_I^0 + \frac{\gamma_0 c^2}{\alpha}\right)} - \frac{c^2}{\alpha}.$$

# IV. Curved generalization of the Möller accelerated frame.

## IV.1. Curved generalization of the Möller accelerated frame which was derived from Einstein-Maxwell field

equations.

It can be shown [8] [9] that the Rindler metric is not a solution to the Einstein field equations for any ponderable source of the gravitational field, i.e., it is a vacuum spacetime [8],[9]. This can be seen directly since the Rindler metric (2.8) can be obtained from Minkowski spacetime via the transformations given by the Eqs.(2.5)-(2.6). The general relativistic analog of the Möller frame was obtained in [7] and [8] The $g_{00}$ component for the uniform gravitation field metric was given by

$$g_{00} = \exp\left(\frac{2a\xi}{c^2}\right). \tag{4.1.1}$$

(in [8] the coordinate was $x$ and $|a|=1$).

$$ds^2 = \exp(2a\xi)d\tau^2 - dx^2 - dy^2 - d\xi^2. \tag{4.1.2}$$

This is a 4D version of the exotic Kaluza-Klein metric which derived by Visser [11] from 5D Einstein-Maxwell field equations:

$$G_{ab} = \lambda g_{ab} + T_{ab},$$

$$T_{ab} = -2\frac{\delta \mathcal{L}}{\delta g^{ab}} + g_{ab}\mathcal{L},$$

$$\mathcal{L} = \frac{1}{4}\mathcal{F}_{ab}\mathcal{F}^{ab},$$

$$A_0 = A(\xi), A_i = 0, i = 1,2,3.$$

$$T_{ab} = \mathcal{F}_{ac}\mathcal{F}^c_b - \frac{1}{4}g_{ab}(\mathcal{F}_{cd}\mathcal{F}^{cd}).$$

$$T^0_0 = -\frac{1}{4}E^2, \lambda = \frac{1}{4}E^2.$$

(4.1.3)

Also noting that $e^{2a\xi} \approx 1 + 2a\xi$ for $a\xi \ll 1$ we might try using $e^{2a|\xi|}$ for $g_{00}$. This same association between the approximate $(1+2a\xi)$ and exact $e^{2a\xi}$ time component of the metric was given originally by Einstein [10]. What, if anything, do we do for $g_{ii}$ where $i = 1,2,3$. If we make the simplest asumption that only $g_{00}$ should be non-trivial we arrive at ($c = 1$):

$$ds^2 = \exp(2a|\xi|)d\tau^2 - dx^2 - dy^2 - d\xi^2. \tag{4.1.4}$$

Again, if one calculates the Einstein tensor $G_{\alpha\beta}$, the only non-zero terms are

$$G_{xx} = G_{yy} = a\delta(\xi) + a^2. \tag{4.1.5}$$

## IV.2. Derivation metric of the curved frame using Born rigidity conditions. The general structural equation for the holonomic Lorentzian manifold. Born rigid constant accelerated motion on a curved Lorentzian manifold.

Born rigidity [12],[13],[14],[15],[16],[17] proposed by and later named after Max Born, is a concept in special relativity. It is one answer to the question of what, in special relativity, corresponds to the rigid body of non-relativistic classical mechanics.

The defining property of Born rigidity is locally constant distance in the co-moving frame for all points of the body in question. It is a very restrictive sense of rigidity; for example, it is impossible to put a disk into rotation while maintaining its Born rigidity. Several weaker substitutes have been proposed as rigidity conditions.

**Notation** *We will argue that in fact we can implement Born's notion of rigid motion in both flat spacetime and arbitrary curved non-holonomic spacetimes containing classical and Colombeau's distributional sources.*

Let us consider holonomic Lorentzian manifold $\mathcal{L} = (\mathbf{M}, g)$. Suppose that $V_\nu(x), x = (x_0, x_1, x_2, x_3, x_4), \nu = 0, 1, 2, 3, 4$ are smooth vector field on $\mathbf{M}$ such that

$$g_{\mu\nu}V^\mu V^\nu = 1.$$

$$\frac{\partial^2 V_\nu}{\partial x^\epsilon \partial x^\sigma} = \frac{\partial^2 V_\nu}{\partial x^\sigma \partial x^\epsilon}. \tag{4.2.1}$$

$$\mu,\nu = 0,1,2,3,4.$$

It is well known that that the next equality is satisfied [18]:

$$\nabla_\mu V_\nu = \Sigma_{\mu\nu} + \Omega_{\mu\nu} + V_\mu F_\nu, \tag{4.2.2}$$

where

$$\Sigma_{\mu\nu} = \nabla_{(\mu} V_{\nu)} - V_{(\mu} F_{\nu)},$$

$$\Omega_{\mu\nu} = \nabla_{[\mu} V_{\nu]} - V_{[\mu} F_{\nu]}, \tag{4.2.3}$$

$$F_\mu = V^\nu \nabla_\nu V_\mu.$$

It is easy to check that the next equality is satisfied:

$$2\nabla_{[\epsilon}\nabla_{\sigma]} V_\nu = 2\partial_{[\epsilon}\partial_{\sigma]} V_\nu + \left( \frac{\partial \Gamma^\mu_{\epsilon\nu}}{\partial x^\sigma} - \frac{\partial \Gamma^\mu_{\sigma\nu}}{\partial x^\epsilon} + \Gamma^\mu_{\sigma\rho}\Gamma^\rho_{\epsilon\nu} - \Gamma^\mu_{\epsilon\rho}\Gamma^\rho_{\sigma\nu} \right) V_\mu. \tag{4.2.4}$$

From Eqs. (4.2.1)-(4.2.4) by simple calculation finally we obtain the general structural equation for the *holonomic* Lorentzian manifold.

$$R^\mu_{\epsilon\sigma,\nu} V_\mu = 2\nabla_{[\epsilon}\Sigma_{\sigma]\nu} + 2\nabla_{[\epsilon}\Omega_{\sigma]\nu} + 2\nabla_{[\epsilon}(V_{\sigma]} F_\nu). \tag{4.2.5}$$

**Definition** 4.2.1. *Vector field $V_\nu(x)$ on Lorentzian manifold $\mathcal{L} = (\mathbf{M},g)$ such that*

$$\Sigma_{\mu\nu} = \nabla_{(\mu} V_{\nu)} - V_{(\mu} F_{\nu)} = 0,$$

$$\Omega_{\mu\nu} = \nabla_{[\mu} V_{\nu]} - V_{[\mu} F_{\nu]} = 0,$$

(4.2.6)

*is called Born rigid vector field without rotation.*

**Definition** *4.2.2. Suppose that for Lorentzian manifold $\breve{\mathcal{L}} = (\mathbf{M}, g)$ are exist vector field $V_\nu(x)$ such that Eqs.(4.2.6) is satisfied. $\breve{\mathcal{L}} = (\mathbf{M}, g)$ is called Born rigid Lorentzian manifold.*

# Born rigid constant acelerated motion on a curved Lorentzian manifold.

Suppose that

$$\Sigma_{\mu\nu} = \Omega_{\mu\nu} = 0,$$

$$g_{\mu\nu} F^\mu F^\nu = -\mathsf{a}_0^2/c^4.$$

(4.2.7)

$$\mu, \nu = 0, 1, 2, 3, 4.$$

Thus from Eq. (2.1.1) one obtain

$$\nabla_\mu V_\nu = V_\mu F_\nu.$$

(4.2.8)

In the comoving coordinates $X = (X^0, X^1, X^2, X^3, X^4)$ one obtain

$$V_k = V^k = 0,$$

$$V^0 = g_{00}{}^{-1/2}, \tag{4.2.9}$$

$$V_0 = g_{00}{}^{1/2}.$$

Take ansatz for the metric $ds^2$ in the next form:

$$ds^2 = D(X^4)(dX^0)^2 - (dX^1)^2 - (dX^2)^2 - (dX^3)^2 - A(X^4)(dX^4)^2, \tag{4.2.10}$$

Thus

$$V_0 = \sqrt{D(X^4)},$$
$$V^0 = \frac{1}{\sqrt{D(X^4)}}. \tag{4.2.11}$$

Using Eqs.(4.2.11) and Eq.(4.2.8) one obtain:

$$A(X^4) = \frac{c^4}{4a_0^2 D^2(X^4)} \left( \frac{dD(X^4)}{dX^4} \right)^2. \tag{4.2.12}$$

Using the replacement

$$dy^4 = A^{1/2}(X^4) dX^4, \quad X^0 = y^0, \quad X^2 = y^2, \quad X^3 = y^3, \tag{4.2.13}$$

finally we obtain

$$ds^2 = \exp\left(\frac{2a_0 y^4}{c^2}\right)(dy^0)^2 - (dy^1)^2 - (dy^2)^2 - (dy^3)^2 - (dy^4)^2. \tag{4.2.14}$$

One may notice that the metric (4.2.14) can be seen as some curved generalization of the flat Rindler metric which is Minkowski spacetime as seen by an observer undergoing constant, linear acceleration:

$$F^4 = \frac{DV^4}{dS} = \frac{dV^4}{dS} + \Gamma^4_{00}(V^0)^2 =$$

$$= \frac{1}{g_{00}} \Gamma^4_{00} = -\frac{g^{44}}{2g_{00}} \frac{\partial g_{00}}{\partial y^4} = \frac{a_0}{c^2}.$$

(4.2.15)

$$R_{\alpha\beta,\gamma\delta} = \frac{1}{2}\left( \frac{\partial^2 g_{\alpha\delta}}{\partial y^\beta \partial y^\gamma} + \frac{\partial^2 g_{\beta\gamma}}{\partial y^\alpha \partial y^\delta} - \frac{\partial^2 g_{\alpha\gamma}}{\partial y^\beta \partial y^\delta} - \frac{\partial^2 g_{\beta\delta}}{\partial y^\alpha \partial y^\gamma} \right) +$$

$$+ g_{\mu\nu}(\Gamma^\mu_{\beta\gamma}\Gamma^\nu_{\alpha\delta} - \Gamma^\mu_{\beta\delta}\Gamma^\nu_{\alpha\gamma}) = g_{\alpha\sigma} R^\sigma_{\cdot\beta,\gamma\delta}$$

(4.2.16)

Thus

$$R_{10,10} = -\frac{1}{2}\left[ \frac{\partial^2 g_{00}}{\partial y^{1^2}} - \frac{1}{2g_{00}}\left( \frac{\partial g_{00}}{\partial y^1} \right)^2 \right] = -\frac{a_0^2}{c^4} \exp\left( \frac{2a_0 y^1}{c^2} \right)$$

(4.2.17)

$$R_{00} = -R_{10,10}, \quad R_{11} = -\frac{a_0^2}{c^4}, \quad R_{10} = 0.$$

$$R = 2\frac{a_0^2}{c^4}.$$

(4.2.18)

Again, if one calculates the Einstein tensor $G_{\alpha\beta}$, the only non-zero terms are

$$G_{22} = G_{33} = \frac{1}{2}R = \frac{a_0{}^2}{c^4}. \qquad (4.2.19)$$

# References

[1] Hsu J. P.,Kleff S. M., Chin. J. Phys. 36, 768 (1998).
[2] Hsu J. P., Einstein's Relativity and Beyond - New Symmetry Approaches,(World Scientific, Singapore,2000), Chapters 21-24.
[3] Møller C., Danske Vid. Sel. Mat-Fys. 20, No. 19 (1943); See also The Theory of Relativity,(Oxford Univ. Press, London, 1952), pp.253-258.
[4] Wu T., Lee Y.C., Int'l. J. Theore. Phys. 5, 307-323 (1972); Ta-You Wu,Theoretical Physics, vol. 4, Theory of Relativity, (Lian Jing Publishing Co.,Taipei,1978), pp. 172-175, and references therein.
[5] Ernst A., Chin. J. Phys. 40, N.6, (2002).

[6] Podosenov S.A.,"Rectilinear Born's hard continuum motion with a constant acceleration in the accompanying tetrade", in: Col. "Problems of the gravitation theory", Theor. and math. phys. series A, issue 1, Moscow: VNIIOFI, 1972, pp. 95-104 (in Russian).
[7] Podosenov S.A.,Space-time structure and fields of bound charges, "Russian Physics Journal" no.10, 1997.
[8] Desloge E.A., "Nonequivalence of a uniformly accelerating reference frame and a frame at rest in a uniform gravitational field", Am.J.Phys.,57, 1121-1125 (1989).
[9] Tilbrook D.,"General coordinatisations of the flat space-time of constant proper-acceleration", Aust. J. Phys. 50, 851-868 (1997)
[10] Einstein A.,"On the influence of gravitation on the propagation of light," Annals of Physics (Germany) 35, 898 (1911).
[11] Visser M.,"An Exotic Class of Kaluza-Klein Models", Phys. Lett. B 159, 22-25 (1985). See also the eprint version: http://arxiv.org/abs/hep-th/9910093v1

[12] Bona C., Rigid-motion conditions in special relativity.Phys. Rev. D27,1243-1247 (1983)

[13] Eriksen E., Mehlen M., Leinaas J.M.,Relativistic Rigid Motion in One Dimension. Phys. Scr. 25 905-910 (1982)

[14] Born M., (1909). "Die Theorie des starren Elektrons in der Kinematik des Relativitäts-Prinzipes". Ann. Phys. Lpz. 30:1.

[15] Ehrenfest P., (1909). "Gleichförmige Rotation starrer Körper und Relativitätstheorie". Phys. Zeitschrift 10: 918.

[16] Planck M., (1910). "Gleichförmige Rotation und Lorentz-Kontraktion". Phys.Zeitschrift 11: 294.

[17] Einstein A., (1911). "Zum Ehrenfesten Paradoxon". Phys. Zeitschrift 12: 509.

[18] Novikov I.D, Frolov V.P.,Physics of Black Holes.Kluwer Academic Publishers:1989 ISBN-13: 9789027726858 ISBN: 902772685X